\documentclass[aps,pre,preprint,superscriptaddress]{revtex4}
\usepackage{graphicx,subfigure}
\usepackage{amsmath,amsfonts}
\begin{document}
\newcommand{\EQ}{Eq.~}
\newcommand{\EQS}{Eqs.~}
\newcommand{\FIG}{Fig.~}
\newcommand{\FIGS}{Figs.~}
\newcommand{\TAB}{Table~}
\newcommand{\SEC}{Sec.~}
\newcommand{\SECS}{Secs.~}
\title{Self-organization of feedforward structure and entrainment in excitatory neural networks with spike-timing-dependent plasticity}

\author{Yuko K. Takahashi}
\affiliation{
Faculty of Engineering,
The University of Tokyo, 7-3-1, Hongo, Bunkyo-ku, Tokyo 113-8656, Japan}
\affiliation{
School of Medicine,
Tokyo Medical and Dental University, 1-5-45, Yushima, Bunkyo-ku, Tokyo 113-8510, Japan}
\author{Hiroshi Kori}
\affiliation{
Division of Advanced Sciences, Ochadai Academic Production,
Ochanomizu University, 2-1-1, Ohtsuka, Bunkyo-ku, Tokyo 112-8610, Japan}
\author{Naoki Masuda}
\affiliation{
Graduate School of Information Science and Technology,
The University of Tokyo, 7-3-1, Hongo, Bunkyo-ku, Tokyo 113-8656, Japan}

\date{\today}

\begin{abstract}
Spike-timing dependent plasticity (STDP) is an organizing principle of biological neural networks. While synchronous firing of neurons is considered to be an important functional block in the brain, how STDP shapes neural networks possibly toward synchrony is not entirely clear. We examine relations between STDP and synchronous firing in spontaneously firing neural populations. Using coupled heterogeneous phase oscillators placed on initial networks, we show numerically that STDP prunes some synapses and promotes formation of a feedforward network. Eventually a pacemaker, which is the neuron with the fastest inherent frequency in our numerical simulations, emerges at the root of the feedforward network. In each oscillatory cycle, a packet of neural activity is propagated from the pacemaker to downstream neurons along layers of the feedforward network. This event occurs above a clear-cut threshold value of the initial synaptic weight. Below the threshold, neurons are self-organized into separate clusters each of which is a feedforward network.
\end{abstract}

\pacs{}
\maketitle

\section{Introduction} \label{sec:intro}
Synchronous firing of neurons has been widely observed and
 is considered to be a neural code that adds to firing rates.
For example, experimental evidence suggests the relevance of synchronous firing in
 stimulus encoding \cite{Laurent2002},
 feature binding \cite{Singer1995,Engel2001},
 and selective attention \cite{Engel2001, Fries0105}.
Collective dynamical states of neurons including synchrony may appear
 as a result of self-organization based on synaptic plasticity.
Modification of synaptic weights (i.e., weights of edges in the network terminology) often occurs
 in a manner sensitive to relative spike timing of presynaptic and postsynaptic neurons,
 which is called spike-timing-dependent plasticity (STDP).
In the commonly found asymmetric STDP, which we consider in this work,
 long-term potentiation (LTP) occurs when presynaptic firing precedes postsynaptic firing
 by tens of milliseconds or less,
 and long-term depression (LTD) occurs in the opposite case \cite{Gerstner96etc}.
The amount of plasticity is larger when the difference in the presynaptic spike time
 and the postsynaptic spike time is smaller \cite{Gerstner96etc}.

The asymmetric STDP reinforces causal pairs of presynaptic and postsynaptic spikes
 and eliminate other pairs.
Based on this property of STDP,
 how STDP may lead to various forms of synchronous firing
 has been studied in both experiments and theory.
Synchronous firing in the sense of simultaneity of spike timing
 can be established in recurrent neural networks
 when the strength of LTP and that of LTD are nearly balanced \cite{Karbowski2002}.
Large-scale numerical simulations suggest that
 reproducible spatiotemporal patterns of spike trains self-organize
 in heterogeneous recurrent neural networks \cite{STDP_synchro_set,Morrison2007}.
Self-organization of clusters of synchronously firing neurons
 that excite each other in a cyclic manner has also been reported \cite{Levy01etc,Cateau2008}.

We previously showed that STDP leads to formation of feedforward networks
 and entrainment when there is a pacemaker in the initial network \cite{Masuda2007}.
We considered random networks of coupled oscillators whose
 synaptic weights change slowly via STDP.
We assumed that the oscillators have a common inherent frequency
 except a single pacemaker whose inherent frequency is larger.
By definition, the rhythm of the pacemaker is not affected by those of other oscillators.
The network generated via STDP is a feedforward network whose root is the pacemaker.
In a final network, a spike packet travels from the pacemaker to the other neurons
 in a laminar manner.
The neurons directly postsynaptic to the pacemaker fire more or less synchronously
 just after the pacemaker does.
These neurons form the first layer.
These neurons induce synchronous firing of the neurons
 directly postsynaptic to them, which define the second layer.
In this fashion, a spike packet starting from the pacemaker
 reaches the most downstream neurons within relatively short time,
 which resembles the phenomenology of the synfire chain \cite{Diesmann99Abeles91}.
Compared to the case of frozen synaptic weights,
 a pacemaker entrains the rest of the network more easily with STDP
 in the meaning that entrainment occurs with smaller initial synaptic weights.

The previous work does not explain how pacemakers emerge.
No matter whether the pacemakers are intrinsic oscillators or network oscillators,
 they pace rhythms of other elements without being crucially affected by other rhythms.
Although some pacemakers may be ``robust" oscillators whose rhythms are insensitive to general input,
 a more natural explanation may be that pacemakers emerge through synaptic plasticity in a neural network
 in which pacemakers are initially absent.
In this case, emergent pacemakers do not have to be robust oscillators;
 their rhythms can change in response to external input.
The emergent network topology makes such neurons pacemakers
 by eliminating incoming synapses.
A neuron would fire with its own rhythm if it is not downstream to any
neuron. This scenario is actually the case for two-neuron networks
\cite{Zhigulin03pre-Nowotny03jns,Masuda2007}. 
Here we are concerned to networks of more
than two neurons.
An associated question is which oscillator may become a pacemaker.

In this work, we numerically investigate 
recurrent networks of coupled phase oscillators subject to STDP.
We show that, when the initial synaptic weights are strong enough,
 STDP indeed yields feedforward networks
 so that downstream neurons are entrained by an emergent pacemaker.
To our numerical evidence, the emergent pacemaker is always the neuron
 with the largest intrinsic frequency.
Below the threshold for entrainment,
 STDP leads to the segregation of the initial neural network into subnetworks of feedforward networks.

\section{Model}
\subsection{Coupled phase oscillators}
We model dynamics of neural networks by $N$ coupled phase oscillators
whose synaptic weights are plastic.  Although a majority of real
neurons fire in the excitable (i.e., fluctuation-driven) regime,
for tractability we use phase oscillators, which fire in an
oscillatory manner. Generally speaking, phase transitions are more
easily and clearly determined in the oscillatory regime than in the
excitable regime.  This is a reason why collective neural dynamics
\cite{phase-osc,higher} including ones associated with STDP
\cite{Karbowski2002,Seliger02,Masuda2007} have been analyzed in the
oscillatory regime actually to give insights into dynamics of neural
networks possibly operating in the excitable regime.  In the
following, we report numerical results for $N=3$ and $N=100$.

The state of neuron $i$ $(1\leq i \leq N)$ is represented by a phase
variable $\phi_i \in [0, 2 \pi)$.  We identify $\phi_i=0$ and
$\phi_i=2\pi$.  When $\phi_i$ crosses $0$ in the positive direction,
neuron $i$ is defined to fire. 
 We denote by $t_j$ and $t_i$ the spike time
of presynaptic and postsynaptic neurons.  If $\phi_i$ crosses 0 in the
positive direction as time advances from $t$ to $t+\Delta t$, we set
$t_i = t+[2\pi-\phi_i(t)]/[2\pi+\phi_i(t+\Delta t)-\phi_i(t)]\Delta
t$.  As the initial condition, we set $\phi_i=0$ $(1\leq i \leq N)$
for $N=3$.  We adopt this artificial initial condition to draw phase
diagrams to systematically understand possible routes to synchrony via
STDP.  For $N=100$, $\phi_i(0)$ is picked randomly and independently
for each $i$ from the uniform density on $[0,2\pi)$.  Neuron $i$ is
endowed with inherent frequency $\omega _i$ so that it fires regularly
at rate $\omega_i/2\pi$ when isolated.  Connectivity between neurons
is unidirectional and weighted, consistent with the properties of
chemical synapses.  The set of edges in a network is denoted by $E$.
In other words, $(j,i)\in E$ if neuron $j$ is presynaptic to neuron
$i$.  Dynamics of the coupled phase oscillators are given by
\begin{equation}
\frac{\mathrm{d}\phi_i}{\mathrm{d}t} = \omega_i + \frac{1}{\langle k \rangle}\sum_{j:(j,i)\in E}g_{ji} \sin(\phi_j - \phi_i)+\sigma\xi_i,
\label{eq:kuramoto}
\end{equation}
 where $\langle k \rangle$ is the average in-degree of neuron $i$,
 $g_{ji}$ is a synaptic weight, and $\xi_i$ represents the standard Gaussian white noise
 independent for different $i$.
As a result of the phase
reduction theory \cite{Kuramotobook},
the coupling term 
in the oscillatory regime 
is generally given by a $2\pi$-periodic function of the
phase difference $\phi_j - \phi_i$ under the assumption of weak coupling. 
This is also the case for pulse coupling, for which 
averaging an original pulse coupling term over one
oscillatory cycle results in a coupling term as a function of
$\phi_j - \phi_i$ \cite{higher}.
Modeling realistic synaptic coupling needs a coupling term
that contains higher harmonics \cite{higher}.
However, our objective in the present paper is not to precisely
describe the neural dynamics
but to clarify general consequences of STDP under the
oscillatory condition. We thus employ 
the simplest coupling term (i.e sinusoidal coupling).

For $N=3$, we set the 
amplitude of the noise $\sigma=0.0071$ so that
the phase transitions are sharp enough
and artificial resonance that is prone to occur when
inherent frequencies satisfy $M_i\omega_i=M_j\omega_j$
for small integers $M_i$ and $M_j$ $(1 \leq i<j \leq N)$ is avoided.
Accordingly,
an independent normal variable with mean 0 and standard deviation
 $\sigma\sqrt{\Delta t}=0.00071$
 is added to each neuron every time step $\Delta t=0.01$;
 we use the Euler-Maruyama integration scheme with unit time $\Delta t$.
To determine the phase transitions
for $N=100$, we do not apply dynamical noise
because, up to our numerical efforts, the numerical results
do not significantly suffer from artificial resonance.
In some other simulations with $N=100$, we add different amplitudes of 
dynamical noise to examine the robustness of the results.

\subsection{STDP}
With STDP, $g_{ji}$ is repeatedly updated depending on spike timing of neuron $j$ and $i$.
Specifically, LTP occurs when a postsynaptic neuron fires slightly after
 a presynaptic neuron does, and LTD occurs in the opposite case \cite{Gerstner96etc}.
We assume that synaptic plasticity operates much more slowly than firing dynamics.
We denote by $A^+$ and $A^-$ the maximum amount of LTP and that of LTD
 incurred by a single STDP event.
Most of previous theoretical work supposes that $A^-$ is somewhat,
 but not too much, larger than $A^+$, to avoid explosion in firing rates
 and to keep neurons firing \cite{Karbowski2002, STDP_synchro_set,Morrison2007,Levy01etc,Cateau2008,Masuda2007}.
Therefore we set $A^+/A^-=0.9$.
How a single spike pair specifically modifies the synaptic weight is under investigation \cite{RubinandGutig,Morrison2007},
 and triplets or higher-order combination of presynaptic and postsynaptic spikes
 rather than a single presynaptic and postsynaptic spike pair may induce STDP \cite{FroemkeandPfister}.
However, we consider the simplest situation in which STDP modifies synaptic weights
 in an additive manner and the amount of STDP is determined by the relative timing of
 a presynaptic and postsynaptic spike pair.
A single synaptic modification $\Delta g_{ji}$ triggered by a spike pair
 is represented by
\begin{equation}
\Delta g_{ji} = \begin{cases}
		A^+ \exp(-\frac{t_j-t_i}{\tau}) & \text{$t_j-t_i < 0$} \\
		-A^- \exp(\frac{t_j-t_i}{\tau}) & \text{$t_j-t_i > 0$}
		\end{cases},
\label{eq:asym}
\end{equation}
 where $\tau$ is the characteristic time scale of the learning window,
 which is known in experiments to be 10-20 ms \cite{Gerstner96etc}.
Given that inherent frequencies of many pyramidal neurons roughly range between 5 and 20 Hz,
 $\tau$ is several times smaller than a characteristic average interspike interval.
Therefore, following \cite{Masuda2007}, we set $\tau = 1/6\times2\pi/\omega$,
 where $\omega$ is a typical value of spike frequency that is used to determine $\omega_i$.
Following our previous work \cite{Masuda2007},
 we set $\omega = 8.1$.
Because learning is slow compared to neural dynamics,
 $A^-$ must be by far smaller than a typical value of $g$.
To satisfy this condition, we set $A^-=0.001$ for $N=3$.
When $N=100$, average in-degree $\langle k \rangle$ is set equal to 10.
This implies that a neuron receives about five to ten times
 more synapses than when $N=3$.
To normalize this factor,
 we set it $A^-=0.0001$ for $N=100$.

We assume that $g_{ji}$ is confined in $[0,g_{max}]$;
 all the synapses are assumed to be excitatory,
 because the asymmetric STDP explained in \SEC\ref{sec:intro}
 has been found mostly in excitatory synapses.
Because dynamical noise is assumed not to be large,
 all the synaptic weights usually develop
 until $g_{ji}$ almost reaches either $g_{max}$ or 0,
 until when we run each simulation run.
Note that, even if $g_{ji}=0$ is reached, $(j,i)$ still belongs to $E$.
The upper limit $g_{max}$ is determined so that
a notion of synchronization that we define in \SEC\ref{sec:measure}
 does not occur when $g_{ji}=g_{max}$, $\forall (j,i)\in E$.
Accordingly, we set $g_{max}=7.5$ and $g_{max}=15$
 for $N=3$ and $N=100$, respectively.

\subsection{Measurement of synchrony} \label{sec:measure}
To obtain the threshold for synchrony in \SECS\ref{sec:small} and
\ref{sec:100 neurons},
we start numerical simulations 
with the initial condition $g_{ji}=g_0$,
 $\forall (j,i)\in E$.
There are various notions of synchrony.
We focus on the possibility of frequency synchrony
 in which neurons fire at the same rate.
In the oscillatory regime, frequency synchrony is commonly achieved in two main ways.
One is when neurons are connected by sufficiently strong mutual coupling.
Then they oscillate at the same rate and with proximate phases.
The other is when some neurons entrain others.
When upstream neurons, which serve as pacemakers,
 entrain downstream neurons so that they are synchronized in frequency,
 synchronous firing in the sense of spike timing may be missing
 due to synaptic delay.
However, neurons located at the same level in the hierarchy relative to the pacemakers
 tend to have close spike timing \cite{Masuda2007,Kori04Kori06}.
We explore possible emergence of such dynamics when pacemakers are initially absent
 in networks.

We quantify the degree of frequency synchrony
 by order parameter $r$ defined by
\begin{equation}
r = \log_{10} \left[\frac{1}{N} \sum_i\left({\tilde{\omega}}_i-\frac{1}{N}\sum_{i^{\prime}} {\tilde{\omega}}_{i^{\prime}}\right)^2\right],
\label{eq:criteria}
\end{equation}
 where $\tilde{\omega_i}=\mathrm{d}\phi_i/\mathrm{d}t$
 is the actual instantaneous frequency of neuron $i$ when coupled to other neurons.
If all the neurons fire exactly at the same rate,
 $r$ would become negative infinity. In the actual frequency
 synchrony,
$r$ takes a large negative value mainly
because of time discretization.
We manually set $r_c=-4$ for $N=3$ and $r_c=-9$ for $N=100$, so that
$r\le r_c$ corresponds to the full frequency synchrony.
The value of $r_c$ for $N=100$ is smaller than for $N=3$ for two reasons.
First, in the numerical simulations determining
the degree of frequency synchrony,
dynamical noise is present for $N=3$ and
 absent for $N=100$.
Second, we are concerned to the frequency synchrony of {\it all} the neurons
 so that $\sum_i(\tilde{\omega}_i-1/N\sum_{i^{\prime}}\tilde{\omega}_{i^{\prime}})^2$
 is small regardless of $N$; we have to normalize the prefactor $1/N$
 in \EQ\eqref{eq:criteria}.

\section{Results}
\subsection{Networks of three neurons}\label{sec:small}
Our goal is to understand dynamics of large neural networks.
As a starting point,
 we examine network evolution and possibility of frequency synchrony
 using small networks, which will help
 us understand dynamics of large networks.
Two-neuron networks were previously analyzed \cite{Masuda2007}.
We need at least three neurons to understand competition between 
different synapses, pruning of synapses, and effects of heterogeneity.
Accordingly, we examine dynamics of different three-neuron networks under STDP.
\subsubsection{Complete graph}\label{sec:Clique}

\begin{figure}
\includegraphics[width=7cm,clip]{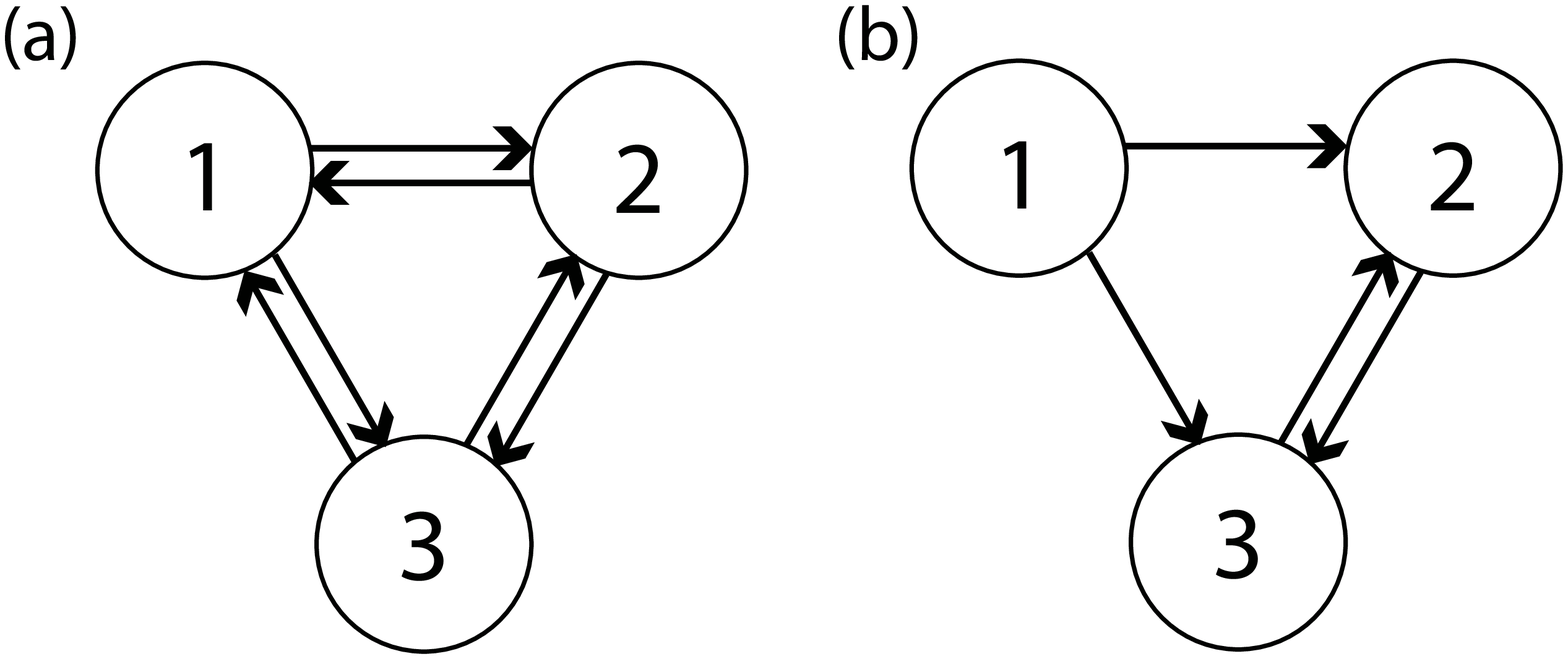}
\caption{Complete graph (a) without a pacemaker
 and (b) with a pacemaker.}
\label{fig:CompleteScheme}
\end{figure}

Consider the complete graph [\FIG{\ref{fig:CompleteScheme}}(a)],
 in which every pair of neurons is bidirectionally connected.
The complete graph does not survive STDP
 because LTP of a synapse implies LTD of the synapse in the reversed direction
 and the amount of LTD is assumed to be larger than that of LTP for the same time lag.
We examine which synapses survive and
 whether frequency synchrony emerges through STDP.
If a predetermined pacemaker exists in a network,
 the activity of the other neurons will be entrained into the rhythm of the pacemaker
 with sufficiently large initial synaptic weights, which was previously shown for $N=2$ and $N=100$ \cite{Masuda2007}.
Here we consider $N=3$ and compare numerical results when a pacemaker is initially present and absent in the complete graph.
Note that the effective initial network
 when the pacemaker neuron 1 is initially present
 is the one shown in \FIG\ref{fig:CompleteScheme}(b),
 because the synapses toward the pacemaker are defined to be
entirely ineffective.

\begin{figure}
\includegraphics[width=12cm,clip]{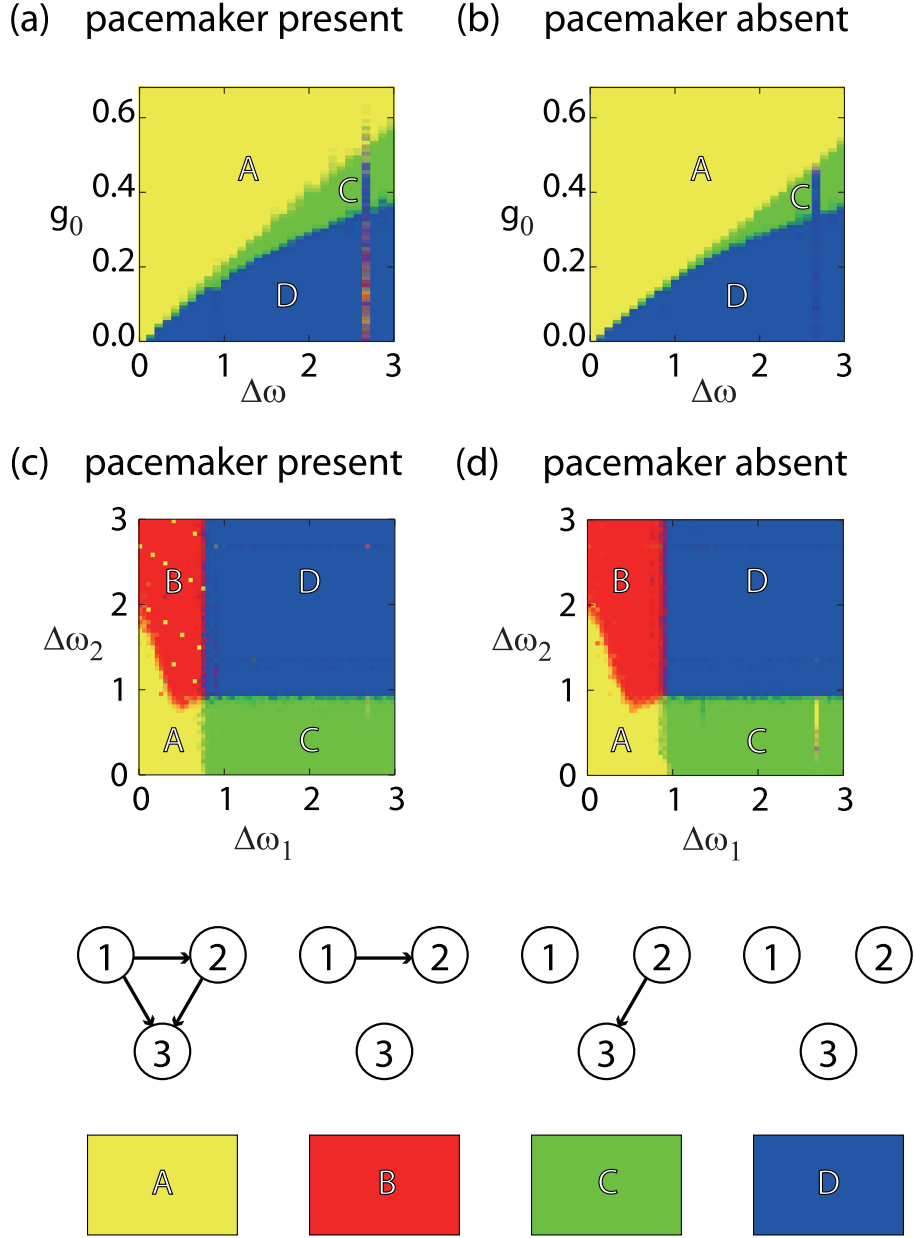}
\caption{(Color) Phase diagrams for the complete graph
 in the [(a) and (b)] $\Delta\omega$-$g_0$
 and [(c) and (d)] $\Delta\omega_1$-$\Delta\omega_2$ spaces.
 One pacemaker neuron is initially present [(a) and (c)] or absent [(b) and (d)].
 We run numerical simulations 20 times for each pair of parameter values.
 We add the red element of the RGB color scheme by the maximum amount divided by 20
 when $g_{12}$ survives in a simulation run.
 Similarly, the green is added when $g_{23}$ survives,
 and the blue is added when all the neurons get disconnected.
 Yellow regions appear when both $g_{12}$ and $g_{23}$ survive,
 since the combination of red and green is yellow.
 In this case, it turns out that $g_{13}$ also survives.
 We verified that no other connectivity,
 such as survival of $g_{13}$ without survival of $g_{12}$ or $g_{23}$,
 and survival of $g_{21}$, $g_{31}$, or $g_{32}$,
 appears except at points near phase transitions and resonance.
 Near phase transitions, we exclude such exceptional
 runs from the statistics.
 In the resonance regions (e.g., $\Delta\omega\approx 2.7$ and $g_0\approx 0.4$),
 the three neurons may remain connected.
 In this situation, however, synaptic weights keep oscillating,
 and any pair of the three neurons is not in frequency synchrony.
 Therefore, we judge such a run as being completely desynchronized and colored blue (labeled D).
}
\label{fig:Complete}
\end{figure}

First, we examine the relation
 between heterogeneity in inherent frequencies, initial synaptic weights, and synchrony.
We expect that small heterogeneity and large initial synaptic weights favor synchrony.
To focus on phase transitions, we reduce the number of parameters
 by setting all the initial synaptic weights equal to $g_0$ and
 restrain inherent frequencies $\omega_1$, $\omega_2$, and $\omega_3$
 ($\omega_1 \geq \omega_2 \geq \omega_3$)
 by imposing, $\omega_1-\omega_2=\omega_2-\omega_3\equiv \Delta \omega$,
where $\omega_2=8.1$.
Numerically obtained phase diagrams are shown in Figs. \ref{fig:Complete}(a) and \ref{fig:Complete}(b)
 for the cases in which a pacemaker is initially present and absent, respectively.
The results are qualitatively the same for the two situations.
The neurons get disconnected and fire independently as a result of STDP
 for sufficiently small $g_0$ or sufficiently large $\Delta \omega$
 (blue regions labeled D).
A feedforward network whose root is the fastest oscillator emerges
 for sufficiently large $g_0$ or sufficiently small $\Delta\omega$ (yellow regions labeled A).
Then all the neurons rotate at frequency $\omega_1$.
In the intermediate regime (green regions labeled C),
 final synaptic weights satisfy $g_{23} \approx g_{max}$
 and $g_{12}$, $g_{13}$, $g_{21}$, $g_{31}$, and $g_{32} \approx 0$.
In this case, neuron 2 entrains neuron 3 so that
 they oscillate at frequency $\omega_2$,
 whereas neuron 1 gets disconnected and oscillates at frequency $\omega_1$.
We rarely observed the case in which neuron 1 entrains 2 (or 3)
 and neuron 3 (or 2) gets isolated.
Although $\omega_1-\omega_2=\omega_2-\omega_3$,
 neuron 1 is more likely to segregate from the network than neuron 3 is.
Quantitatively speaking, Figs. \ref{fig:Complete}(a) and \ref{fig:Complete}(b) indicate that
 the entrainment of the entire network by the fastest neuron (i.e., neuron 1)
 is to some extent easier to realize when the pacemaker is initially absent
 than present (yellow regions labeled A).
In Figs. \ref{fig:Complete}(a) and \ref{fig:Complete}(b),
 the phase diagrams are disturbed along vertical lines at $\Delta\omega\approx2.7$.
This artifact comes from the fact that
 $\omega_1$, $\omega_2$, and $\omega_3$ approximately 
satisfy the resonance condition
 (i.e., $M_1 \omega_1=M_2 \omega_2=M_3 \omega_3$ with small integers $M_1$, $M_2$, and $M_3$).
In some of the following figures,
 similar disturbance appears along special lines.
We can wash away these artifacts by increasing the amount of dynamical noise.
However, we prefer not doing so to prevent the boundaries between different phases
 from being blurred too much.

Next, to examine what happens when $\omega_1$, $\omega_2$, and $\omega_3$
 change independently,
 we set $g_0=0.15$, $\omega_2=8.1$,
 and vary $\Delta \omega_1 \equiv \omega_1-\omega_2$ and
 $\Delta\omega_2 \equiv \omega_2-\omega_3$.
Numerical results with and without a pacemaker
 are shown in Figs. \ref{fig:Complete}(c) and \ref{fig:Complete}(d), respectively.
Figures \ref{fig:Complete}(c) and \ref{fig:Complete}(d) are similar to each other,
 except yellow spots in the red region (labeled B) in \FIG\ref{fig:Complete}(c).
These spots represent entrainment facilitated
 due to the artificial resonance condition satisfied
 by $\omega_1$, $\omega_2$, and $\omega_3$.
Both in Figs. \ref{fig:Complete}(c) and \ref{fig:Complete}(d),
 $g_{23}$ is easier to survive than $g_{12}$ is,
 consistent with Figs. \ref{fig:Complete}(a) and \ref{fig:Complete}(b).
This is indicated by the fact that
 the phase of the frequency synchrony of the three neurons (yellow regions labeled A)
 extends to a larger value of $\Delta\omega_2>0$ along the line $\Delta\omega_1=0$
 than to the value of $\Delta\omega_1>0$ along the line $\Delta\omega_2=0$,
 and that the phase in which neuron 2 entrains 3 (green, C) survives
 up to a larger value of $\Delta\omega_2$ than the value of $\Delta\omega_1$
 up to which neuron 1 entrains neuron 2 but not neuron 3 (red, B).

\begin{figure}
\includegraphics[width=7cm, clip]{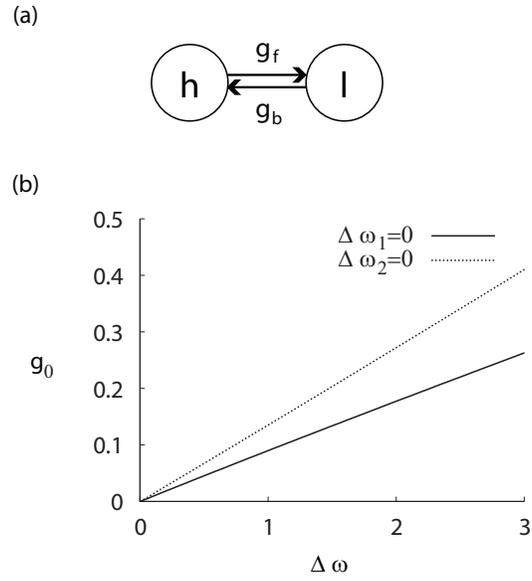}
\caption{(a) Two-neuron network.
 (b) Threshold for frequency synchrony for the two-neuron networks
 corresponding to the $\Delta\omega_1=0$ line and the $\Delta\omega_2=0$ line
 in Figs. \ref{fig:Complete}(c) and \ref{fig:Complete}(d).}
\label{fig:twobodies}
\end{figure}

To examine the cause of the asymmetry in Figs. \ref{fig:Complete}(c) and \ref{fig:Complete}(d)
 along the two lines $\Delta\omega_1=0$ and $\Delta\omega_2=0$,
 we analyze a two-neuron network with asymmetric initial synaptic weights shown in \FIG\ref{fig:twobodies}(a).
The two neurons $h$ and $l$ have inherent frequency $\omega_h$ and $\omega_l$ ($\leq\omega_h$).
The weights of the synapse from neuron $h$ to neuron $l$
 and that from neuron $l$ to neuron $h$
 are denoted by $g_f$ and $g_b$, respectively.
When $\Delta \omega_1=0$ and $\Delta\omega_2\geq 0$ in the three-neuron network,
 neurons 1 and 2 are synchronized almost from the beginning,
 in both frequency and phase,
 because $\omega_1=\omega_2$.
This is true if a trivial condition $g_{12}+g_{21}>0$ is satisfied.
Then the network is reduced to the two-neuron network
 by identifying $\omega_h=\omega_1=\omega_2$, $\omega_l=\omega_3$,
 $g_f=g_{13}+g_{23}$, and $g_b=(g_{21}+g_{31})/2$.
When, $\Delta\omega_1\geq 0$ and $\Delta\omega_2=0$ in the three-neuron network,
 neurons 2 and 3 are synchronized in frequency and phase
 as far as $g_{23}+g_{32}>0$.
Then the network is reduced to the two-neuron network
 with $\omega_h=\omega_1$, $\omega_l=\omega_2=\omega_3$,
 $g_f=(g_{12}+g_{13})/2$, and $g_b=g_{21}+g_{31}$.
For these two situations, we calculate the
 threshold for frequency synchrony in the two-neuron network
 using the semi-analytical method developed in \cite{Masuda2007}.
Because all the synaptic weights are initially equal to $g_0$ in \FIG\ref{fig:Complete},
 the initial condition for the two-neuron network is
 $(g_f,g_b)=(2g_0,g_0)$ for $\Delta\omega_1=0$, $\Delta\omega_2\equiv\Delta\omega\geq 0$,
 and $(g_f,g_b)=(g_0,2g_0)$ for $\Delta\omega_1\equiv\Delta\omega\geq 0$, $\Delta\omega_2=0$.
The phase-transition curves for the frequency synchrony are
 shown in \FIG\ref{fig:twobodies}(b), indicating that
 the threshold is larger along the $\Delta\omega_2=0$ line
 than along the $\Delta\omega_1=0$ line.
This is consistent with the three-neuron results shown in Figs. \ref{fig:Complete}(c) and \ref{fig:Complete}(d).

\subsubsection{Feedforward loop}
\begin{figure}
\includegraphics[width=12cm,clip]{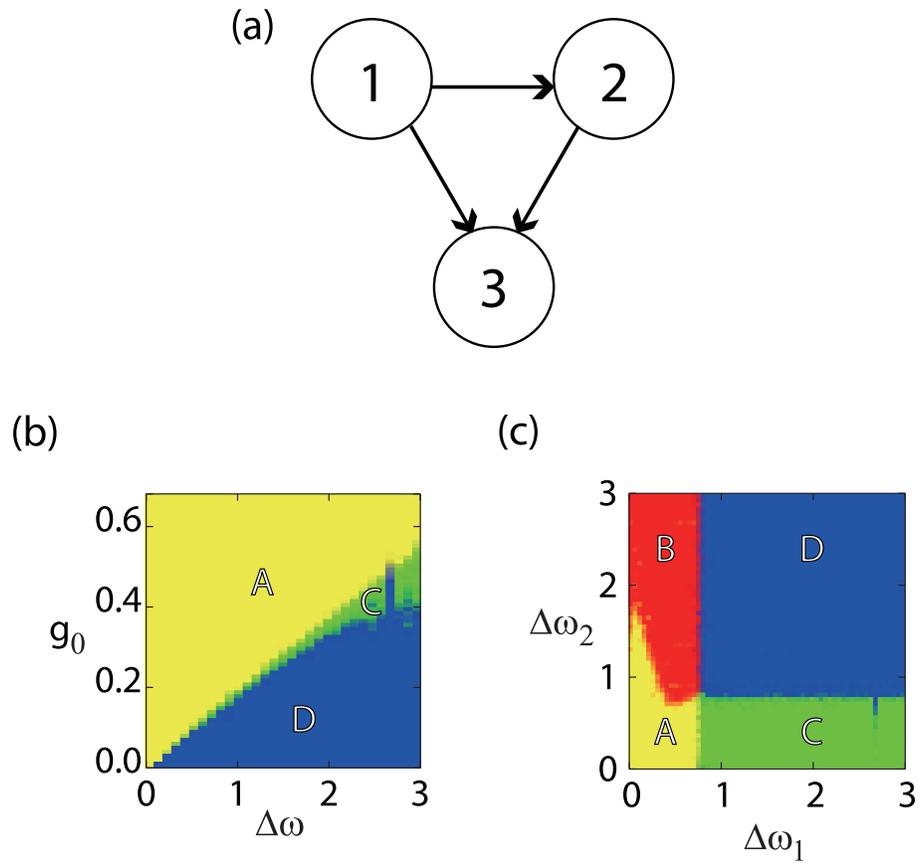}
\caption{(Color) (a) Feedforward loop.
 [(b) and (c)] Phase diagrams for the feedforward loop in two different parameter spaces.
 See \FIG\ref{fig:Complete} for the color code.}
\label{fig:FFL}
\end{figure}

Other three-neuron networks, particularly feedforward ones, are
 presumably embedded in larger neural networks
 in the course of network evolution.
First, we consider the network shown \FIG\ref{fig:FFL}(a)
 as the initial network.

Figure \ref{fig:FFL}(a) is the phase diagram
 in which we vary $\Delta\omega=\omega_1-\omega_2=\omega_2-\omega_3$
 and $g_0=g_{12}=g_{13}=g_{23}$.
The original network shown in \FIG \ref{fig:FFL}(a) survives STDP
 when initial synaptic weights are large or the heterogeneity is small (yellow region labeled A).
In the opposite situation, all the neurons get disconnected
 and fire independently (blue, D).
Neuron 1 detaches from the network and neuron 2 entrains neuron 3
 in the intermediate regime (green, C).

The phase diagram in the $\Delta\omega_1$-$\Delta\omega_2$ parameter space with
 $g_0=0.15$ is shown in \FIG\ref{fig:FFL}(b), which looks similar to Figs. \ref{fig:Complete}(c) and \ref{fig:Complete}(d).
As in the case of the complete graph,
 the situation in which neuron 1 entrains neuron 2 with neuron 3 isolated
 is less likely to arise
 than that in which neuron 2 entrains neuron 3 with neuron 1 isolated.

\subsubsection{Fan-in network}
\begin{figure}
\includegraphics[width=9cm, clip]{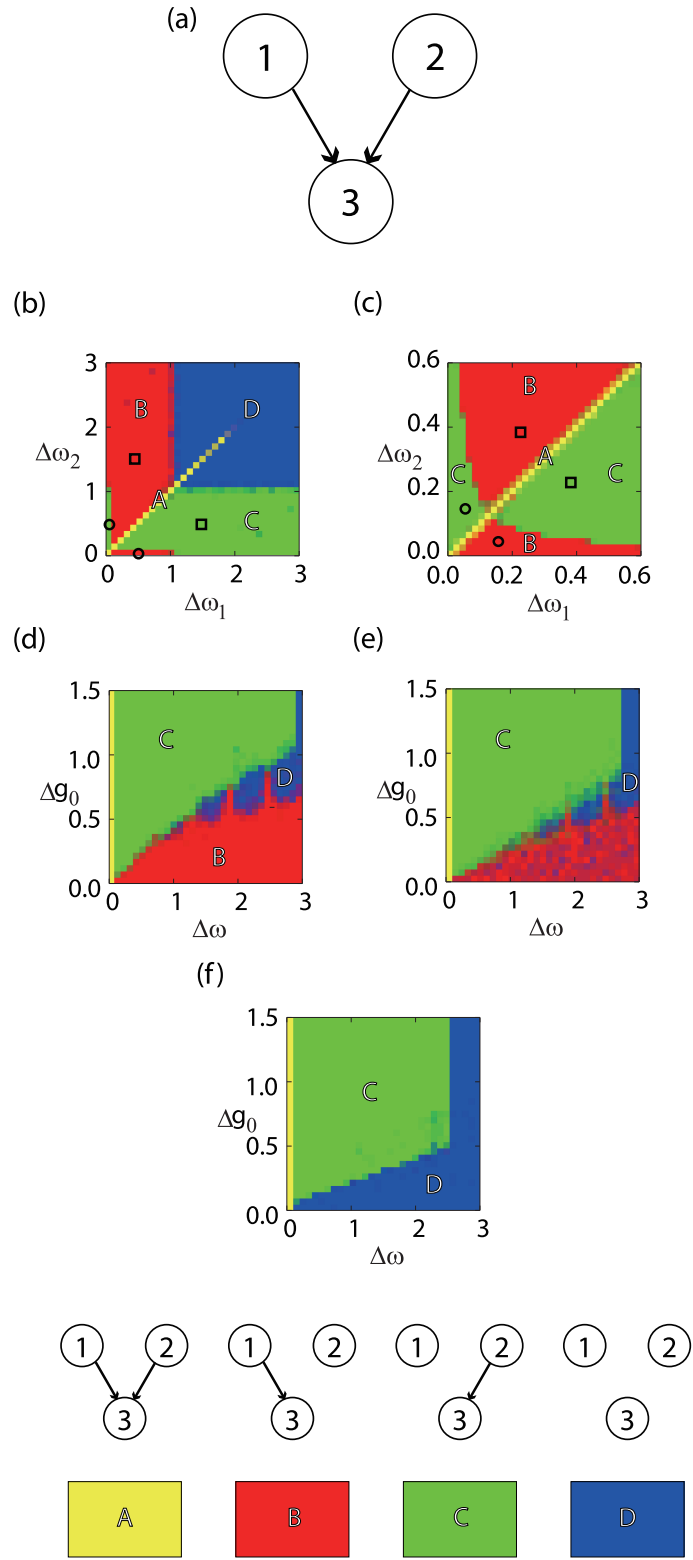}
\caption{(Color) (a) Fan-in network.
 [(b) and (c)] Phase diagrams for the fan-in network
 in the $\Delta \omega_1$-$\Delta \omega_2$ space, with
 (c) being an enlargement of (b). We set $g_0=0.2$.
 [(d), (e), and (f)] Phase diagrams
 in the $\Delta \omega$-$\Delta g_0$ space.
 We set  $g_0=0.2$, (d) $\omega_1=\omega_3+0.8$,
 (e) $\omega_1=\omega_3+1.0$, and (f) $\omega_1=\omega_3+1.2$.}
\label{fig:FI}
\end{figure}

Next, we examine dynamics starting from the fan-in network shown in \FIG \ref{fig:FI}(a).
In this network, neuron 3 is postsynaptic to two pacemaker neurons 1 and 2.
We are concerned to which neuron entrains neuron 3.

First, we examine the case in which two synapses are initially equally strong
 and the inherent frequencies of the two upstream neurons are different.
Accordingly we set $g_{13}=g_{23}=g_0$, $\omega_1-\omega_3\equiv\Delta\omega_1$,
 $\omega_2-\omega_3\equiv\Delta\omega_2$, $g_0=0.2$, and $\omega_3=8.1$.
Figures \ref{fig:FI}(b) and \ref{fig:FI}(c) are the phase diagrams in the $\Delta\omega_1$-$\Delta\omega_2$ space,
 with \FIG\ref{fig:FI}(c) being an enlargement of \FIG\ref{fig:FI}(b).
There are principally four phases:
 neither neuron 1 or 2 entrains neuron 3 (blue regions labeled D),
 both neurons 1 and 2 entrain neuron 3 (yellow, A),
 only neuron 1 entrains neuron 3 (red, B),
 and only neuron 2 entrains neuron 3 (green, C).
The phase diagram is symmetric with respect to the diagonal line $\Delta\omega_1=\Delta\omega_2$.
When $\omega_1$ and $\omega_2$ are too far from $\omega_3$,
 all the neurons get disconnected (blue, D).
Both $g_{13}$ and $g_{23}$ survive only when $\omega_1\approx\omega_2$ (yellow, A).
This phase extends to the disconnection phase (blue, D) on the diagonal
 because, on this line, the firing of neuron 1 elicits LTP of both synapses
 so does firing of neuron 2.
However, this situation is not generic in that $\omega_1$ and $\omega_2$
 must be very close for this to happen.
When $\omega_1$ and $\omega_2$ are not close to each other and not too far from $\omega_3$,
 which upstream neuron entrains neuron 3 is not obvious.
Figure \ref{fig:FI}(b) tells that a necessary condition for an upstream neuron
 to entrain neuron 3 is that the difference between its inherent frequency and $\omega_3$
 is less than $\approx 1.0$.
This condition roughly corresponds to the requirement for the entrainment
 in the two-neuron feedforward network with $g_0=0.2$.
This explains the two rectangular regions $\Delta\omega_1>1.0$, $\Delta\omega_2<1.0$,
 and $\Delta\omega_1<1.0$, $\Delta\omega_2>1.0$ of \FIG\ref{fig:FI}(b).
In the remaining region (i.e. $\Delta\omega_1<1.0$ and $\Delta\omega_2<1.0$),
 the upstream neuron whose inherent frequency is closer to $\omega_3$,
 equivalently, the slower upstream neuron, largely wins the competition
 (regions marked by $\Box$).
The faster upstream neuron entrains neuron 3
 when the inherent frequency of the slower upstream neuron is very close to $\omega_3$
 (regions marked by $\bigcirc$).
The total size of the latter regions is much smaller than
 that of the former regions.

Starting with asymmetric synaptic weights, that is, $g_{13}\neq g_{23}$,
 the upstream neuron more strongly connected to neuron 3
 may entrain neuron 3.
To investigate the interplay of this effect and heterogeneity in the inherent frequency,
 we perform another set of numerical simulations with
 $\omega_1=\omega_3+1$, $\omega_2=\omega_1+\Delta\omega$,
 $g_{13}=g_0$, and $g_{23}=g_0+\Delta g_0$.
The asymmetry in the initial synaptic weight is parameterized by $\Delta g_0$.
Figures \ref{fig:FI}(d)-\ref{fig:FI}(f) shows the phase diagrams in the $\Delta \omega$-$\Delta g_0$
 space for three different values of $\omega_1$.
On the singular line $\Delta\omega=0$ (i.e., $\omega_1=\omega_2$), $\Delta g_0\geq 0$,
 both upstream neurons entrain neuron 3.
On the line $\Delta\omega\geq 0$ (i.e., $\omega_1<\omega_2$), $\Delta g_0=0$, neuron 1,
 whose inherent frequency $\omega_1$ is closer to $\omega_3$ than $\omega_2$ is,
 entrains neuron 3 if $\omega_1$ is not too apart from $\omega_3$ [\FIG\ref{fig:FI}(d)].
This is consistent with the results in Figs. \ref{fig:FI}(b) and \ref{fig:FI}(c).
However, if $g_{23}$ is sufficiently larger than $g_{13}$,
 neuron 2 overcomes the disadvantageous situation $\omega_2-\omega_3>\omega_1-\omega_3$
 to win against neuron 1 and entrains neuron 3.
We confirmed that neuron 2 exclusively entrains neuron 3
 when $\Delta\omega<0$ and $\Delta g_0>0$ (not shown).

\subsection{Networks of many neurons}
In this section, we use networks of heterogeneous $N=100$ neurons to examine
 what network structure and dynamics self-organize
 via STDP when we start from random neural networks.
The inherent frequencies of the neurons are
 independently picked from the truncated Gaussian distribution with mean 8.1, standard deviation 0.5,
 and support $\omega_i\in[7.6,8.6]$.
We assume that every neuron has $\langle k \rangle =10$
 randomly selected presynaptic neurons on average
 so that an arbitrary pair of neurons is connected by a directed edge with probability
 $\langle k \rangle / (N-1) \approx 0.1$.
Except in \SEC\ref{sub:hetero}, where we investigate effects of heterogeneity, the initial synaptic weight is assumed to be $g_0$ common for all the synapses.
We vary $g_0$ as a control parameter.

\subsubsection{Threshold for frequency synchrony}\label{sec:100 neurons}
We compare how STDP affects the possibility of entrainment and formation of
 feedforward networks when a pacemaker is present and when absent.
To this end, we fix a random network and a realization of $\omega_i$ ($1\leq i \leq N$).
Without loss of generality, we assume $\omega_1\geq \omega_2\geq \cdots \geq \omega_N$.
For the network with a pacemaker, we make the fastest neuron a pacemaker.
By definition, the rhythm of the pacemaker is not affected by those of the other neurons
 even though the pacemaker is postsynaptic to approximately $\langle k \rangle$ neurons.
Using the bisection method, we determine the threshold value of $g_0$
 above which all the neurons will synchronize in frequency.

The results without dynamical noise (i.e., $\sigma=0$)
are summarized in \TAB I.
When the pacemaker is present from the beginning,
 STDP drastically reduces the threshold for entrainment \cite{Masuda2007}.
After entrainment, all the neurons rotate at the inherent frequency of the pacemaker,
 that is, $\omega_1=8.60$.
When a pacemaker is initially absent,
 STDP reduces the threshold for frequency synchrony by $34 \%$.
Facilitation of frequency synchrony in the absence of the initial pacemaker
 is consistent with the results for the complete graph with $N=3$ (\FIG\ref{fig:Complete}).
In this situation,
 the scenario to frequency synchrony is different between the presence and the absence of STDP.
With STDP, the fastest oscillator eventually entrains the entire network
 when the initial synaptic weight is above the threshold,
 as in the case of the network with a prescribed pacemaker.
Without STDP, the fastest oscillator does not entrain the other neurons.
The realized mean frequency 8.08 is close to the mean inherent frequency of the 100 neurons.
This suggests that frequency synchrony in this case is achieved by mutual interaction,
 rather than by one-way interaction underlying the entrainment by the fastest neuron.
Therefore, in networks without predetermined pacemakers,
 STDP enables emergence of pacemakers and changes the collective dynamics drastically.

\begin{table}
\caption{Comparison of the threshold for frequency synchrony $g_c$
 and the actual mean frequency of the neurons $\left<\tilde{\omega}\right>$
 in the frequency synchrony. We calculated $\left<\tilde{\omega}\right>$
 by averaging the instantaneous frequency over all the neurons and
 over the last ten unit times of the simulation.}
\begin{tabular}{|c|c|c|c|}
\hline
\multicolumn{2}{|c|}{}&\multicolumn{2}{|c|}{Pacemaker}\\ \cline{3-4}
\multicolumn{2}{|c|}{}&Present&Absent\\ \hline
&Present&$g_c=9.8$&$g_c=0.72$\\
STDP&&$\langle\tilde{\omega}\rangle=8.60$&$\langle\tilde{\omega}\rangle=8.60$\\ \cline{2-4}
&Absent&$g_c=51$&$g_c=0.93$\\
&&$\langle\tilde{\omega}\rangle=8.60$&$\langle\tilde{\omega}\rangle=8.08$\\ \hline
\end{tabular}
\end{table}

\subsubsection{Network dynamics}\label{sub:100 dynamics}

\begin{figure}
\includegraphics[width=14cm,clip]{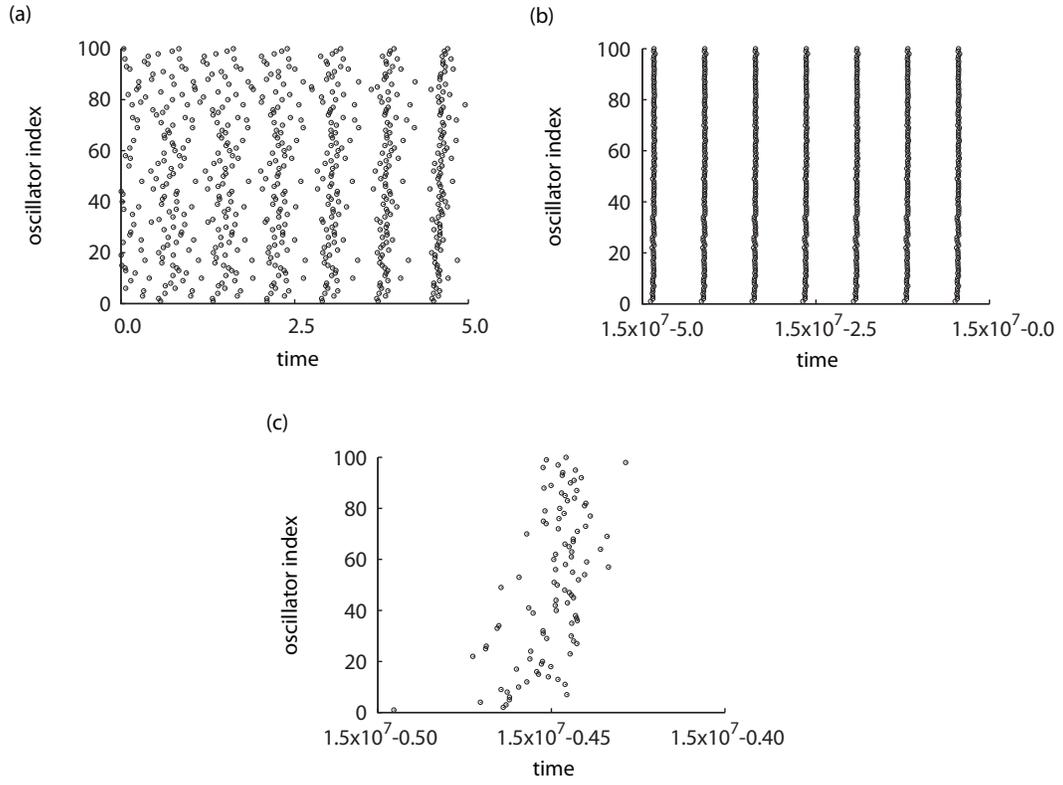}
\caption{Rastergrams for (a) initial 5 time units and
 (b) final 5 time units of a simulation run.
(c) is an enlargement of (b).
 We set $N=100$, $g_0=1.0$, and $\sigma=0$.
 The neurons are aligned according to the order of the inherent frequency.
 }
\label{fig:rastergram}
\end{figure}

\begin{figure}
\includegraphics[width=7cm,clip]{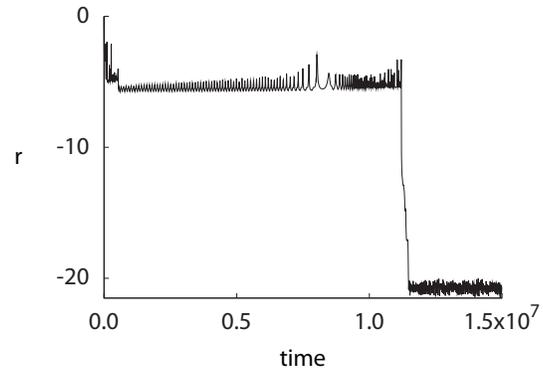}
\caption{Time course of the degree of synchrony when $N=100$,
$g_0=1.0$, and $\sigma=0$.
 The values of $r$ are plotted every 10000 time units.}
\label{fig:measurement}
\end{figure}

\begin{figure}
\includegraphics[width=7cm,clip]{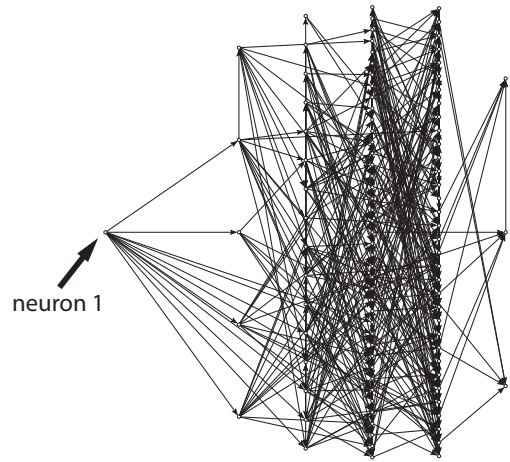}
\caption{Final network structure when $N=100$, $g_0=1.0$,
and $\sigma=0$. The network is drawn by Pajek \cite{pajek}.}
\label{fig:network}
\end{figure}

For $\sigma=0$, example rastergrams when there is initially no
pacemaker and $g_0=1.0$, which is above the threshold value 0.72 (see
\TAB I), are shown in \FIG\ref{fig:rastergram}.  Figures
\ref{fig:rastergram}(a) and \ref{fig:rastergram}(b) correspond to the
initial and final stages of a simulation run under STDP, respectively;
frequency synchrony appears as a result of STDP.  Figure
\ref{fig:rastergram}(c), which is an enlargement of
\FIG\ref{fig:rastergram}(b), shows that the fastest neuron entrains
the other neurons and that faster neurons tend to fire earlier in a
cycle.  Figure \ref{fig:measurement} shows the time course of the
degree of synchrony $r$. Around $t=1.2$ x $10^7$, $r$ sharply drops,
and all the neurons start to oscillate at the same frequency.  The
effective network defined by the surviving synapses in the final state
is drawn in \FIG\ref{fig:network}.  The neurons are placed so that the
horizontal position represents relative spike time in a cycle. With
this ordering, the neurons form a feedforward network.  In other
words, after STDP,
if a presynaptic neuron fires later than a postsynaptic neuron
in a cycle, this synapse is not present.

\begin{figure}
\includegraphics[angle=-90,width=7cm,clip]{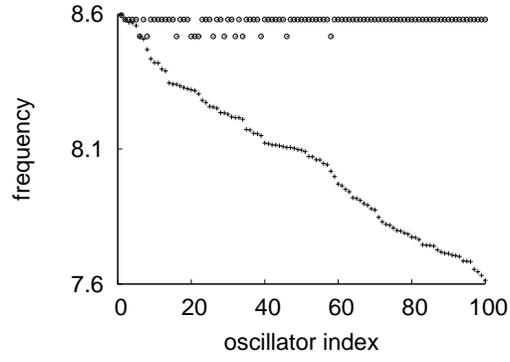}
\caption{Segregation into clusters when $N=100$, $g_0=0.5$, and
$\sigma=0$.
 Inherent frequencies ($+$) and actual frequencies after STDP ($\circ$)
 are shown. We estimate the actual frequencies from the
phase shifts with bins of width 10 time units.}
\label{fig:cluster}
\end{figure}

Partial entrainment occurs when $g_0$ is slightly or moderately smaller than
 the threshold value 0.72.
Circles and crosses in \FIG\ref{fig:cluster} represent the actual frequency after transient
 and the inherent frequency of the each neuron, respectively, when $g_0=0.5$.
The neurons with the same actual frequency belong to the same cluster.
Each cluster forms a feedforward network emanating from an emergent pacemaker.
Figure \ref{fig:cluster} indicates that the neurons are divided into two clusters and one isolated neuron.
Neuron 2 entrains 85 other neurons all of which are slower than neuron 2,
 neuron 6 entrains 12 slower neurons,
 and neuron 1 gets isolated.
In this and further numerical simulations we performed,
 the root of a feedforward subnetwork is always occupied
 by the fastest neuron in the cluster.

Whether two neurons eventually belong to the same cluster is determined
 by where these neurons are located on the initial random network
 and by how close their inherent frequencies are.
If $g_0$ is smaller than the value used for \FIG\ref{fig:cluster},
 two neurons have to be closer in $\omega_i$ to stay connected after STDP.
Then the number of clusters increases, and the number of neurons
 in a cluster decreases on average.

\subsubsection{Robustness against dynamical noise and heterogeneity}\label{sub:hetero}

To examine the robustness of the numerical results reported in
\SEC\ref{sub:100 dynamics}, we perform additional numerical simulations with
dynamical noise and random initial synaptic weights. We draw initial
$g_{ji}$ $(j,i)\in E$ randomly and independently 
from the uniform density on $[0, 2g_0]$, where $g_0=1.0$.

\begin{figure}
\includegraphics[width=14cm,clip]{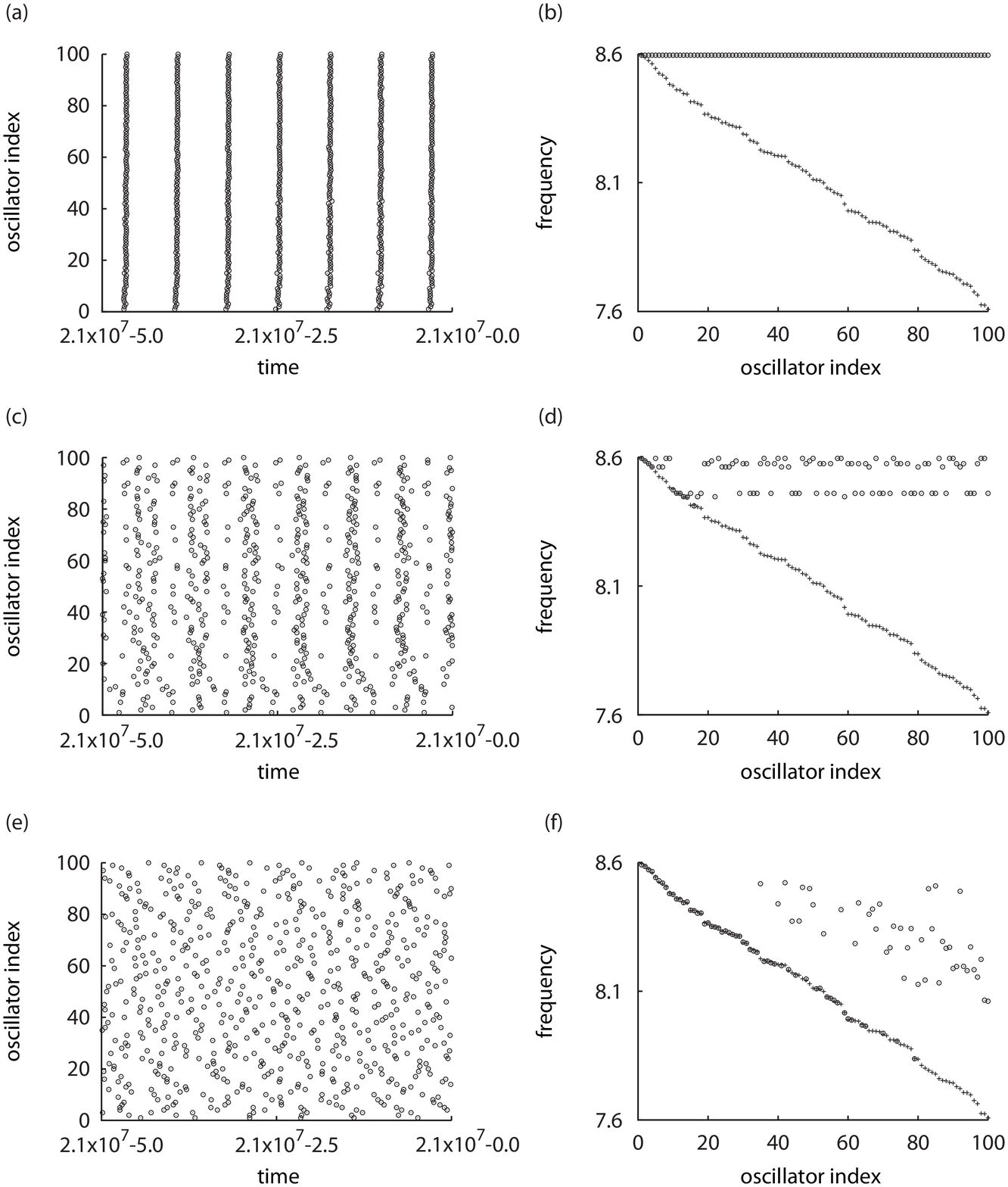}
\caption{Results for $N=100$, $g_0=1.0$, with dynamical noise and heterogeneity in the initial synaptic weights.
We set [(a) and (b)] $\sigma=0.081$, [(c) and (d)] $\sigma=0.405$, and [(e) and (f)]
$\sigma=0.81$. Rastergrams for 5 time units after transient are shown in
(a), (c), and (e).
The neurons are aligned according to the order of the inherent frequency.
Inherent frequencies ($+$) and actual frequencies ($\circ$)
are shown in (b), (d), and (f). Because of the 
dynamical noise, we estimate the actual frequencies from the
phase shifts with bins of width $10^5$ time units.}
\label{fig:N=100 sigma large}
\end{figure}

With
$\sigma=0.081$, the rastergram and
the actual frequency of the neurons after transient are shown in
Figs. \ref{fig:N=100 sigma large}(a) and \ref{fig:N=100 sigma large}(b),
respectively.
With $\sigma=0.081$,
the standard deviation of the accumulated noise in a unit time, which
is equal to $\sigma$, corresponds to 1\% of the phase advancement estimated
by the mean inherent frequency of the oscillators,
which is equal to 8.1. The rastergram
[\FIG\ref{fig:N=100 sigma large}(a)] is indicative of
full entrainment. Indeed,
all the neurons eventually rotate at the inherent
frequency of the fastest neuron [\FIG\ref{fig:N=100 sigma large}(b)].
With $\sigma=0.405$, the neurons are divided into six synchronous
clusters 
of size 31, 26, 19, 12, 5, 4, plus
three isolated neurons
[Figs. \ref{fig:N=100 sigma large}(c) and \ref{fig:N=100 sigma large}(d)].
 With $\sigma=0.81$, many neurons,
particularly faster ones, rotate at their inherent
frequencies [Figs. \ref{fig:N=100 sigma large}(e) and \ref{fig:N=100 sigma large}(f)].
Consequently, there are many clusters of neurons. The frequency synchrony within
each cluster is blurred by dynamical noise.

In sum, emergence of entrainment via STDP survives 
some dynamical noise and heterogeneity in the initial synaptic
weights. We have confirmed that, when the entrainment occurs, it is quickly established
at around $t=10^6$, and the fastest oscillator is located at the
root of the feedforward network, as in \FIG\ref{fig:network}.

\subsubsection{Network motifs}
We investigated the evolution of three-neuron networks in \SEC\ref{sec:small}
 because we expect that these results have something common with
 evolution of such subnetworks in large networks.
The results in \SEC\ref{sec:small} predict the following:
\begin{itemize}
\item Bidirectional edges do not survive STDP,
 and feedforward networks  of size three will be relatively abundant after STDP.
Subnetworks abundant in a large network
 relative to the case of random networks with the same mean degree (or other order parameters)
 are called network motifs \cite{Milo2002}.
The hypothesis that feedforward networks are motifs in large neural networks
 is consistent with the
 observations in {\it C. elegans} neural networks \cite{Milo2002}.

\item As a result of STDP, a neuron has at most one effective upstream neuron
 unless multiple upstream neurons are very close in frequency.
\end{itemize}

There are 13 connected network patterns of three nodes.
How often each pattern appears in a network with $N=100$,
 relative to the random network, can be quantified by the $Z$ score \cite{Milo2002}.
The $Z$ score is the normalized number of a pattern in the network,
 where normalization is given by the mean and the standard deviation of the count of the pattern
 based on independent samples of the randomized networks.
A pattern with a large $Z$ score is a motif of the network with $N=100$.

\begin{figure}
\includegraphics[width=7cm,clip]{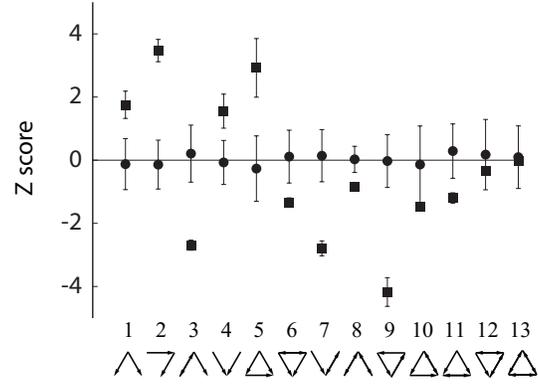}
\caption{Normalized abundance of different three-neuron network patterns.
 We set $N=100$, $g_0=5.0$, and $\sigma=0$. 
Circles and squares correspond to the initial and
 final stages of the simulation runs, respectively.} 
\label{fig:motif}
\end{figure}

Figure \ref{fig:motif} shows the $Z$ score of each pattern before (circle) and after (square) STDP,
 calculated by m-finder \cite{Alonhp}.
We set $\sigma=0$ (i.e., no dynamical noise) in this analysis.
The error bar shows a range of one standard deviation based on ten simulation runs
 in each of which we draw a different initial random network and a different realization
 of $\omega_i$ ($1\leq i\leq 100$).
Before STDP, the neural network is a directed random graph, so that the $Z$ score
 for each pattern is around zero,
 meaning that no pattern is overrepresented or underrepresented significantly.
After STDP, the feedforward network whose emergence and survival were observed
 in \SEC\ref{sec:small} (i.e., pattern 5 in \FIG\ref{fig:motif})
 and patterns consistent with this (i.e., patterns 1 and 2) are overrepresented.
These are motifs of our final networks.
Pattern 4 is also a motif in spite of our negation in \SEC\ref{sec:small}
 because the two upstream neurons in pattern 4 have the same actual frequency.
They are generally different in inherent frequency but
 share a more upstream ancestor.
As the example network in \FIG\ref{fig:network} shows,
 existence of multiple paths from a neuron to another due to
 branching and uniting of edges is compatible with STDP.
The other network patterns are not significant or underrepresented.
These results are further evidence that feedforward networks are formed by STDP
 in heterogeneous neural networks.

\section{Discussion}
We have shown using heterogeneous coupled phase oscillators that
feedforward networks spontaneously emerge via STDP when the initial
synaptic weights are above the threshold value. When this is the
case, the pacemaker emerges at the root of the feedforward network and
entrains the others to oscillate at the inherent frequency of the
pacemaker. Although these results have been known for two-neuron
  networks \cite{Zhigulin03pre-Nowotny03jns,Masuda2007}, we have shown
them for the cases of three and more neurons and quantified the phase
transitions separating frequency synchrony and asynchrony.
 The route to frequency synchrony is distinct from a
conventional route to frequency synchrony that occurs when mutual, but
not one-way, coupling between oscillators is strong enough.  Some
results obtained in this work are unique to the networks without a
prescribed pacemaker.  First, the emergent pacemaker is the fastest
oscillator neuron according to our extensive numerical
simulations. Note that all the oscillators fire at this frequency in
the entrained state, whereas they fire at the mean inherent frequency
of the oscillators when the frequency synchrony is realized by strong
mutual coupling in the absence of STDP.  Second, when the initial
coupling strength is subthreshold, the neurons are segregated into
clusters of feedforward networks.  Third, our numerical evidence
suggests that entrainment under STDP occurs more easily when a
prescribed pacemaker is absent than present.

In spite of a wealth of evidence that real neural circuits are full of
recurrent connectivity \cite{DouglasandWilson}, feedforward structure
may be embedded in recurrent neural networks for reliable
transmission of information
\cite{Diesmann99Abeles91,Klemm05}. Feedforward transmission of
synchronous volleys in rather homogeneous neural networks as those used
in this work serves as a basis of reproducible transmission of more
complex spatiotemporal spike patterns in more heterogeneous networks.
Such patterns may code external input or appear as a result of neural
computation \cite{Diesmann99Abeles91,STDP_synchro_set}.  Feedforward
structure is also a viable mechanism for traveling waves often found
in the brain \cite{Ermentrout01neuron}. Although computational
  roles of feedforward network structure are not sufficiently identified,
our results give
a support to the biological relevance of feedforward networks. The formation of feedforward networks, which we have shown for oscillatory neurons,
is consistent with numerical results for more realistic
excitable neurons subject to STDP \cite{Song01}. The neurons that
directly receive external input may be more excited and fire at a
higher rate compared to other parts of a neural circuit.  Our results
suggest that such a neuron or an ensemble of neurons is capable of
recruiting other neurons into entrainment and creating feedforward
structure.

We assumed the additive STDP with the nearest-neighbor rule
 in which the dependence of the amount of plasticity on the current synaptic weight
 and the effects of distant presynaptic and postsynaptic spike pairs, triplets,
 and so on, are neglected.
Generally speaking, evolution of synaptic weights is affected by
 the implementation of the STDP rule \cite{RubinandGutig}.
However, we believe that our results are robust in the variation of the STDP rule
 as far as it respects the enhancement of causal relationships
 between presynaptic and postsynaptic pairs of neurons.
Our preliminary numerical data with excitable neuron models suggest that
 the results are similar between the multiplicative rule \cite{RubinandGutig}
 and the additive rule [Kato and Ikeguchi (private communication)].
Recent reports claim the relevance of acausal spike pairs
 in the presence of synaptic delay \cite{Cho2007,Cateau2008}.
This and other factors, such as different time scales of LTP and LTD \cite{Song01},
 may let bidirectional synapses survive as observed in {\it in vitro} experiments
 \cite{Song2005}.
Incorporating these factors is an important future problem.

We have ignored inhibitory neurons for two reasons.
First, our main goal is to identify phase transitions regarding
synchrony with a simple model. Second, specific 
rules of STDP are not established for inhibitory neurons, albeit some
pioneering results \cite{Woodin-Haas}. Taking inhibitory neurons 
into account, preferably in the subthreshold regime, warrants for future
work.

\begin{acknowledgments}
We thank Hideyuki C\^{a}teau for valuable discussions.
N.M acknowledges the support from the Grants-in-Aid for Scientific Research
 (Grants No. 20760258 and No. 20540382) and the Grant-in-Aid for Scientific Research on
 Priority Areas: Integrative Brain Research (Grant No. 20019012) from MEXT, Japan.
H.K acknowledges the financial support from the Grants-in-Aid for Scientific
 Research (Grant No. 19800001) from MEXT, Japan, and Sumitomo Foundation (Grant No. 071019).
\end{acknowledgments}

\newpage

\clearpage
\pagestyle{empty}

\clearpage

\clearpage

\clearpage

\clearpage

\clearpage

\clearpage

\clearpage
    
\clearpage

\clearpage
\end{document}